\documentclass[reprint,pra,superscriptaddress,amsmath,amssymb,aps,longbibliography,]{revtex4-1}

\usepackage{graphicx}% Include figure files
\usepackage{dcolumn}% Align table columns on decimal point
\usepackage{bm}% bold math
\usepackage{color}
\usepackage{blindtext}
\usepackage{appendix}
\usepackage{lipsum}
\usepackage{xr}
\usepackage{xcolor}
\usepackage{siunitx}
\usepackage{multirow}
\usepackage[colorlinks]{hyperref}% add hypertext capabilities
\hypersetup{%
	plainpages=true,
	breaklinks=true,% not default in dvips mode, so we must specify
	hypertexnames=false,%not ideal, but needed when pagenums duplicate (`i' vs. `1')
	pageanchor=true,
	colorlinks=true,
	linkcolor={blue},
	citecolor={blue},
	urlcolor={blue},
	anchorcolor={black}
}

\newcommand{\bt}{\textbf}

\begin{document}

\title{Flux-noise-resilient transmon qubit via a doubly-connected gradiometric design}

\author{J.~B.~Fu} 
\affiliation{College of Computer Science and Technology, National University of Defense Technology, Changsha 410073, China}
\author{Da-Wei Wang}
\email{dw-wang@mail.tsinghua.edu.cn}
\affiliation{School of Integrated Circuits, Tsinghua University, Beijing 100084, China}
\author{B.~Ren} 
\affiliation{College of Computer Science and Technology, National University of Defense Technology, Changsha 410073, China}
\author{Z.~H.~Yang} 
\affiliation{College of Computer Science and Technology, National University of Defense Technology, Changsha 410073, China}
\author{S.~Hu}
\affiliation{College of Computer Science and Technology, National University of Defense Technology, Changsha 410073, China}
\author{G.~Y.~Huang} 
\affiliation{College of Computer Science and Technology, National University of Defense Technology, Changsha 410073, China}
\author{S.~H.~Cao}
\affiliation{College of Computer Science and Technology, National University of Defense Technology, Changsha 410073, China}
\author{D.~D.~Liu} 
\affiliation{College of Computer Science and Technology, National University of Defense Technology, Changsha 410073, China}
\author{X.~F.~Zhang} 
\affiliation{College of Computer Science and Technology, National University of Defense Technology, Changsha 410073, China}
\author{X.~Fu}
\affiliation{College of Computer Science and Technology, National University of Defense Technology, Changsha 410073, China}
\author{S.~C.~Xue} 
\affiliation{College of Computer Science and Technology, National University of Defense Technology, Changsha 410073, China}
\author{Y.~G.~Che}
\affiliation{College of Computer Science and Technology, National University of Defense Technology, Changsha 410073, China}
\author{Yu-xi Liu}
\affiliation{School of Integrated Circuits, Tsinghua University, Beijing 100084, China}
\affiliation{Frontier Science Center for Quantum Information, Beijing 100084, China}
\author{M.~T.~Deng} 
\email{mtdeng@nudt.edu.cn}
\affiliation{College of Computer Science and Technology, National University of Defense Technology, Changsha 410073, China}
\author{J.~J.~Wu}
\affiliation{College of Computer Science and Technology, National University of Defense Technology, Changsha 410073, China}

% \date{\today}

\begin{abstract}
Frequency-tunable superconducting transmon qubits are a cornerstone of scalable quantum processors, yet their performance is often degraded by sensitivity to low-frequency flux noise. Here we present a doubly-connected gradiometric transmon (the ``8-mon") that incorporates a nano-airbridge to link its two loops. This design preserves full electrical tunability and remains fully compatible with standard X-mon control and readout, requiring no additional measurement overhead. The airbridge interconnect eliminates dielectric loss, which enables the 8-mon to achieve both energy relaxation times $T_{\rm 1}$ comparable to reference X-mons and, in the small flux-bias regime, a nearly threefold enhancement in Ramsey coherence time $T_{\rm 2}^*$. This improved $T_{\rm 2}^*$ reaches the same order as $T_{\rm 1}$ without employing echo decoupling. The device also exhibits superior long-term frequency stability even without any magnetic field shielding. We develop a spatially correlated flux-noise model whose simulations quantitatively reproduce the experimental coherence trends, revealing the coexistence of short- and long-correlation-length magnetic noise in the superconducting chip environment. By unifying high tunability with intrinsic flux-noise suppression through a robust geometric design, the 8-mon provides a practical pathway toward more coherent and stable superconducting quantum processors.
\end{abstract}

\maketitle
\section*{Introduction}
The pursuit of large-scale fault-tolerant quantum computation has established superconducting circuits as one of the leading physical platforms~\cite{Mohseni2025, Croot2025,Gu2017}. Among them, the transmon qubit~\cite{Koch2007} has become the central building block, owing to its robustness against charge noise and millisecond-scale coherence times~\cite{Bland2025}. Transmon-based quantum processors have achieved quantum computational advantage~\cite{Arute2019, Wu2021, Morvan2024, Gao2025}, demonstrated error correction below the fault-tolerance threshold~\cite{Acharya2025, He2025}, and are now advancing toward practical quantum advantage~\cite{Kim2023,Abanin2025}. Among various designs, the frequency-tunable transmon retains distinct advantages. Its tunability affords precise individual qubit addressability and enables high-fidelity gates such as the parametric control-phase gate~\cite{Arute2019, Wu2021, Li2025, Jin2025}. Furthermore, it allows for the dynamic spectral avoidance of coherent interactions with microscopic two-level system (TLS) defects in amorphous materials, thereby enhancing both operational fidelity and device yield.

\begin{figure*}[t]
\includegraphics[width=0.9\linewidth]{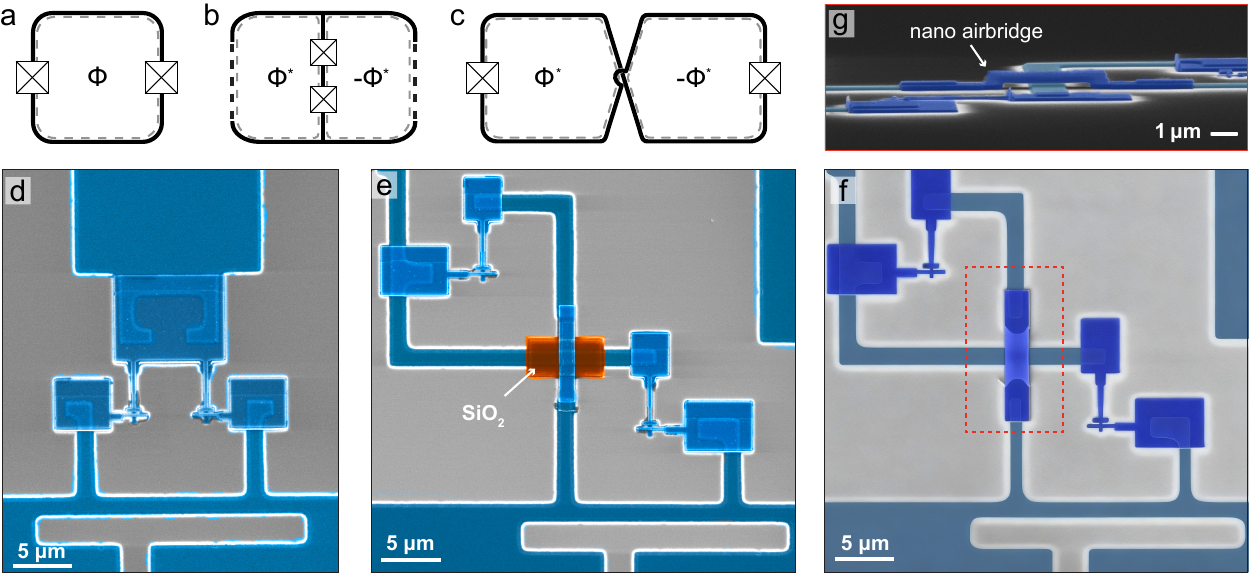}
\caption{\bt{Device architectures of frequency-tunable superconducting qubits.} \bt{a}-\bt{c}, Schematic illustrations of a conventional SQUID (\bt{a}), a triply-connected gradiometric SQUID (\bt{b}), and a doubly-connected gradiometric SQUID (\bt{c}). In \bt{b}, dashed lines indicate circuit segments that may vary across different types of superconducting qubits. In \bt{c}, a cross-bridge connects the two loops, forming the gradiometric structure. \bt{d}, Pseudo-colored scanning electron microscopy (SEM) image of a reference X-mon qubit, identical to the devices measured in this work. \bt{e}, SEM image of an 8-mon qubit implemented with an embedded SiO$_2$ dielectric layer to bridge the two superconducting loops. \bt{f}, An alternative 8-mon design utilizing a nano-airbridge for inter-loop connection (outlined by the red dashed area). \bt{g}, A zoom-in SEM image of the nano-airbridge highlighted in \bt{f}.}
\label{fig1}
\end{figure*}

Frequency-tunable transmons commonly utilize a superconducting quantum interference device (SQUID) to enable transition frequency modulation via an external magnetic flux (Fig.~\ref{fig1}\bt{a}). This design, however, makes the qubit coherence acutely sensitive to external magnetic flux noise~\cite{PhysRevLett.99.187006}. Fluctuations in the ambient field (originating from material-based TLSs~\cite{RevModPhys.86.361, Gusenkova2022, Rower2023} or other external sources~\cite{Wu2025}) couple directly into the SQUID loop, becoming a significant source of dephasing and consequently degrading the stability of the qubit frequency. Existing strategies to mitigate flux noise often involve optimizing material purity and magnetic field shielding, yet the intrinsic sensitivity of the SQUID geometry to flux-noise remains a fundamental limitation. 

%%%%%%%%%%%%%%%%%%%%%% FIG. 1 %%%%%%%%%%%%%%%%%%%%%%%%
%%%%%%%%%%%%%%%%%%%%%%%%%%%%%%%%%%%%%%%%%%%%%%%%%%%

Gradiometric designs offer a route to balance functional tunability and flux-noise-induced decoherence. Planar gradiometric qubits based on triply-connected loops (Fig.~\ref{fig1}\bt{b}) have been implemented in flux qubits~\cite{Koch2006, Paauw2009, Schwarz2013, Kim2022, Berlitz2025}, fluxonium qubits~\cite{Gusenkova2022, Benatre2025}, and some superconducting-spin qubits~\cite{Pita2023}. However, this architecture is not directly transferable to transmon qubits. Furthermore, even in these implementations, imperfect cancellation of magnetic flux noise occurs due to unequal supercurrent distribution between the two loops, which limits the efficacy of common-mode rejection.

In this work, we design a novel transmon qubit that incorporates a symmetrically counter-wound, doubly-connected gradiometric SQUID loop, resembling a figure ``8" with a bridging junction at the loop crossing (Fig.~\ref{fig1}\bt{c}). We term this architecture the \textit{8‑mon} qubit, which is coined in analogy to the ``X‑mon"~\cite{Barends2013} and consistent with earlier reports of ``eight‑shaped gradiometric" designs~\cite{Paauw2009, Schwarz2013}. This topology ensures that the net magnetic flux coupled to the qubit is proportional to the spatial gradient of the magnetic field across the loop, rather than to its absolute magnitude~\cite{Huber2008}. As a result, the 8-mon qubit becomes largely immune to quasi-static or low-frequency magnetic noise while retaining full electrical tunability. Moreover, the device exhibits excellent long‑term frequency stability.

\section*{Results}
\subsection{Qubit structure, fabrication and measurement}
We have fabricated and characterized three separate chips (denoted as chip-\emph{a}, \emph{b}, and \emph{c}), each containing multiple 8-mon devices alongside reference X-mons for comparative analysis. For clarity, we denote the chip identifier in the subscript of the qubit name, and use an asterisk (*) in the superscript to specifically label an 8-mon qubit. For instance, qubit $\rm Q_{\rm a2}^*$ refers to an 8-mon device on chip-\emph{a}, whereas $\rm Q_{\rm c1}$ indicates a standard X-mon qubit on chip-\emph{c}.

Both 8-mon and X-mon qubits feature individual XY-drive and flux-bias lines (similar to the device layout in Ref.~\cite{Zhang2025}). For each 8-mon, the flux line is positioned to couple preferentially to only one of the two loops. To ensure the gradiometric cancellation of uniform magnetic fields, the two SQUID arms and the two Josephson junctions of the 8-mon were designed to be geometrically identical, thereby minimizing the disparity in their effective loop areas. SEM images of the studied X-mon and 8-mon devices are illustrated in Figs.~\ref{fig1}\bt{d}-\ref{fig1}\bt{e}. The combined area of the two loops in each 8-mon is set to twice that of the reference X‑mon loop.

%%%%%%%%%%%%%%%%%%%%%% FIG. 2 %%%%%%%%%%%%%%%%%%%%%%%%
\begin{figure}[ht]
\includegraphics[width=\linewidth]{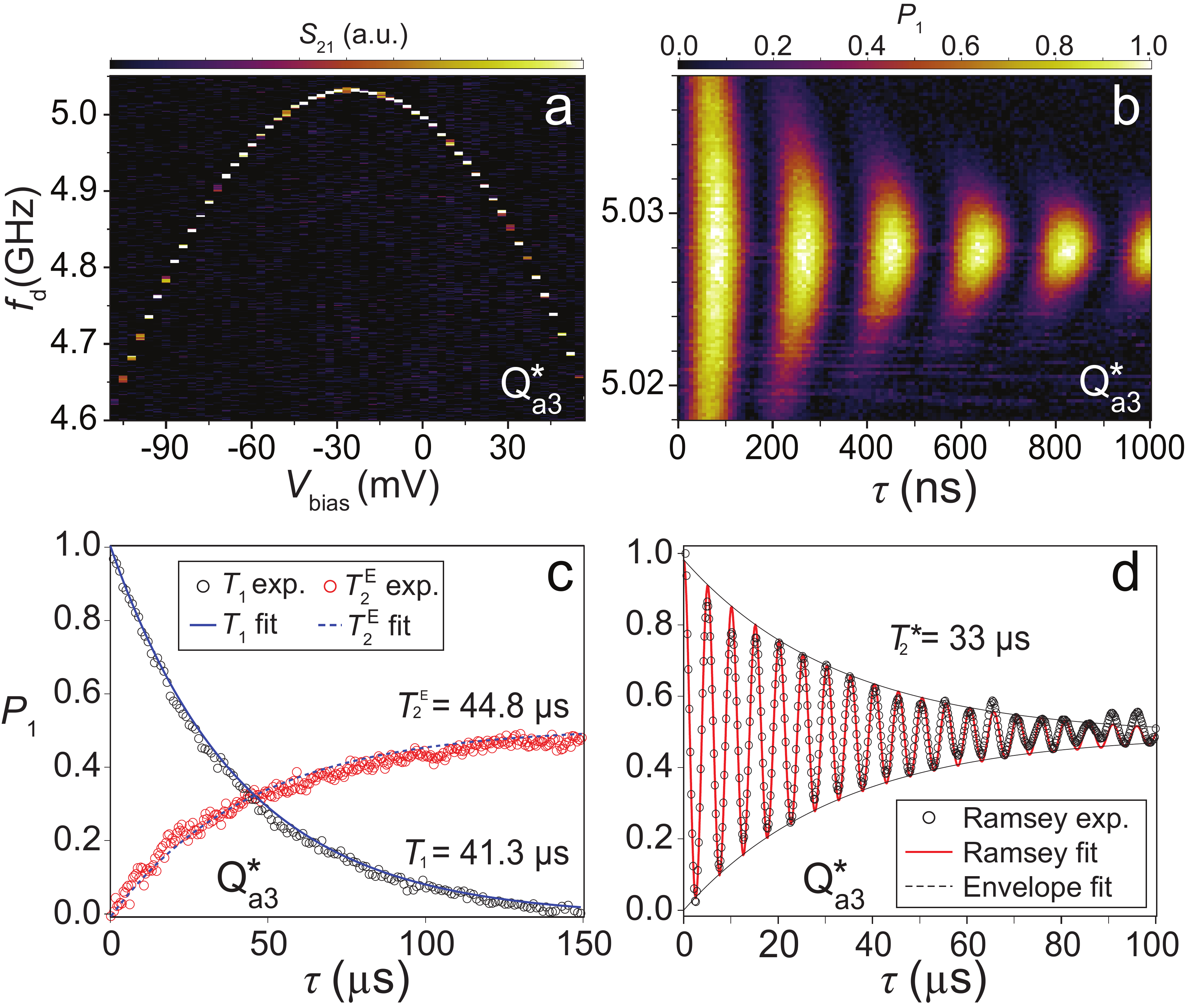}
\caption{\bt{Basic characterization of an 8-mon qubit.} \bt{a}, Qubit transition frequency as a function of the bias voltage applied to the flux line, demonstrating that the 8-mon qubit remains broadly tunable, akin to a conventional transmon. \bt{b}, Rabi oscillations, demonstrating high-fidelity coherent control of the qubit state. \bt{c}, Energy relaxation and echo decoherence measurements, revealing good coherence times with exponential fits yielding $T_{\rm 1}=41.3$ $\mu$s and $T_{\rm 2}^{\rm E}=44.8$ $\mu$s. \bt{d}, Ramsey oscillation fringe, from which a Ramsey coherence time $T_{\rm 2}^*\approx 33$ $\mu$s is extracted. All data in panels \bt{b-d} were acquired at the flux-insensitive sweet spot (optimal point). Measurements were performed on device $\rm Q_{\rm a3}^*$.}
\label{fig2}
\end{figure}
%%%%%%%%%%%%%%%%%%%%%%%%%%%%%%%%%%%%%%%%%%%%%%%%%%%

 The 8-mon and reference X-mon devices were fabricated on c-plane sapphire substrates using 100-nm-thick aluminum films, largely following established processes for X-mons. Josephson junctions were formed by double-angle evaporation in a Dolan-bridge geometry~\cite{Dolan1977}. A key distinction for the 8-mon is the implementation of the cross-bridge that connects the two loops of the gradiometric SQUID. We developed two distinct fabrication methods for this structure. The first employs a deposited SiO$_2$ dielectric layer to isolate the bottom superconducting arm from the top arm, similar to the approach used for a spin qubit in Ref.~\cite{Pita2023}. The second method utilizes a nano-airbridge~\cite{DengPatent2025}, thereby avoiding any dielectric layer. In this study, chip-\emph{a} was fabricated with the nano-airbridge design, while chip-\emph{b} and chip-\emph{c} employed the SiO$_2$ dielectric approach.

The operation and readout of the 8-mon qubit follow the same procedure as for the X‑mon. All measurements were conducted in a dilution refrigerator with a base temperature of approximately 15 mK. To investigate the impact of magnetic environmental noise, chip-\emph{a} and chip-\emph{b} were measured with a cryogenic magnetic shield (fabricated from $\mu$-metal), while chip-\emph{c} was deliberately measured without one. Further details on the measurement setup are provided in the Supplementary Information (~\ref{Setup}~\&~\ref{Readout}).

%%%%%%%%%%%%%%%%%%%%%% FIG. 3 %%%%%%%%%%%%%%%%%%%%%%%%
\begin{figure*}[ht]
\includegraphics[width=\linewidth]{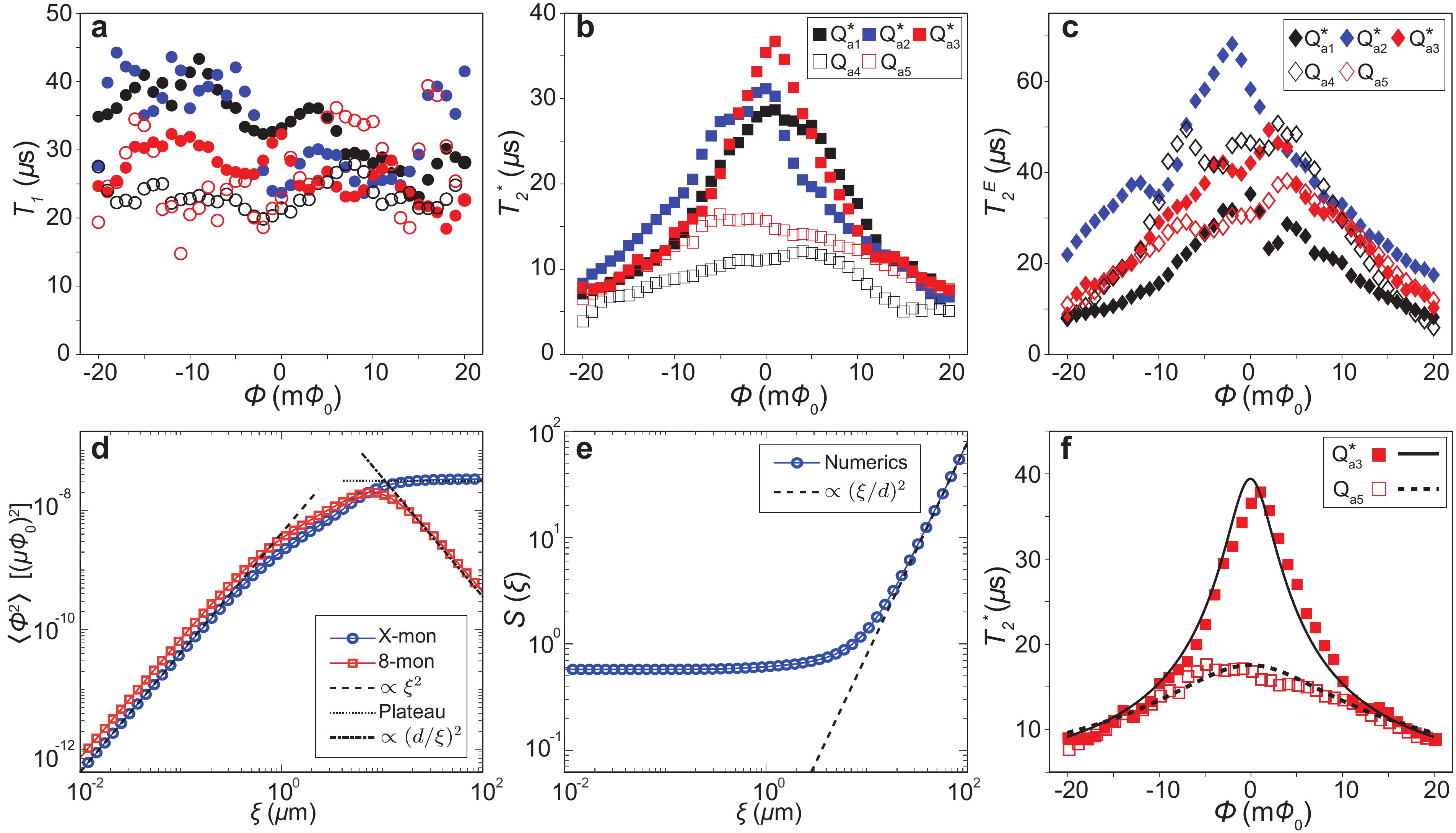}
\caption{\bt{Coherence time characterization.} \bt{a}, Energy relaxation time $T_{\rm 1}$ for three 8-mon qubits and two reference X-mon qubits from the same chip, measured near the flux-insensitive sweet spot. The $T_{\rm 1}$ values, ranging from 20 $\mu$s to 45 $\mu$s, indicate that the 8-mon design preserves a similar relaxation time to the conventional transmon. \bt{b}, Ramsey coherence time $T_{\rm 2}^*$ under small applied flux bias. The 8-mon qubits exhibit a substantial enhancement, with $T_{\rm 2}^*$ values nearly 2–3 times longer than those of the X-mons near the optimal point. This improvement diminishes as the flux bias increases. \bt{c}, Echo decoherence time $T_{\rm 2}^{\rm E}$ within the same bias regime. The $T_{\rm 2}^{\rm E}$ values for both 8-mon and X-mon qubits are of the same order of magnitude.  \bt{d}, Flux-noise variance $\langle\Phi^2\rangle$ as a function of the spatial
magnetization correlation length $\xi$ for X-mon and 8-mon circuits, calculated
using the continuous magnetization model.
\bt{e}, Noise-suppression factor
$S(\xi)=\langle\Phi_X^2\rangle/\langle\Phi_8^2\rangle$ as a function of $\xi$.
\bt{f}, Experimental measurements of the Ramsey dephasing time $T_2^*$ as a
function of flux bias $\Phi_{\rm bias}$ for two representative devices
(Q$^*_{a3}$ and Q$_{a5}$).
The solid and dashed lines show fits obtained using the same dephasing model. }
\label{fig3}
\end{figure*}
%%%%%%%%%%%%%%%%%%%%%% FIG. 3 %%%%%%%%%%%%%%%%%%%%%%%%

\subsection{Characterization of qubit coherence}

Using device $\rm Q^*_{\rm a3}$ as a representative example, we perform basic characterization of the 8‑mon qubit. The transition frequency shifts smoothly with the bias current applied to its flux line (Fig.~\ref{fig2}\bt{a}), showing a tunability characteristic similar to that of a conventional frequency‑tunable X‑mon~\cite{Barends2013}. The flux responsivity of the 8‑mon is comparable to reference X‑mons (Supplementary Information), indicating that the magnetic field gradient generated by its dedicated flux line is sufficient for effective control.

We then applied resonant microwave pulses through the XY drive line to coherently control the qubit state. Clear Rabi oscillations were observed (Fig.~\ref{fig2}\bt{b}), demonstrating successful excitation and manipulation of the 8-mon. Subsequently, we characterized the coherence properties at the flux-insensitive sweet spot. The energy relaxation time $T_{\rm 1}$, Hahn echo coherence time $T_{\rm 2}^{\rm E}$, and Ramsey coherence time $T_{\rm 2}^*$ were extracted from exponential fits to the measured decay curves (Figs.~\ref{fig2}\bt{c} and \ref{fig2}\bt{d}), yielding $T_{\rm 1}$=41.3 $\mu$s, $T_{\rm 2}^{\rm E}$=44.8 $\mu$s, and $T_{\rm 2}^{\rm *}$=33 $\mu$s, respectively. These results indicate excellent coherence properties of the 8-mon qubit.

To comprehensively investigate the coherence times of the 8-mon under flux bias in comparison with reference X-mons, we applied a small-range flux bias and recorded the corresponding changes in $T_{\rm 1}$, $T_{\rm 2}^*$, and \(T_2^\text{E}\). The characterization was primarily conducted on devices from chip-\emph{a}. 

 As shown in Fig.~\ref{fig3}\bt{a}, 8‑mons and X‑mons show comparable $T_{\rm 1}$ values, primarily distributed between 20–45~$\mu$s at small flux bias, confirming that the 8‑mon topology itself does not degrade the relaxation time. However, the specific cross‑bridge fabrication process strongly influences $T_{\rm 1}$. Compared to the nano‑airbridge used in chip‑\emph{a}, 8‑mons on chip‑\emph{b} with a SiO$_2$ dielectric bridge exhibit a markedly shorter $T_{\rm 1}$ of only $\sim$2~$\mu$s (Supplementary Information \ref{S4}). This reduction is likely attributable to a high density of defects introduced by the amorphous SiO$_2$ layer~\cite{PhysRevLett.95.210503}.

The phase coherence of the 8-mon qubits was then characterized. As shown in Fig.~\ref{fig3}\bt{b}, under small flux bias the Ramsey coherence time $T_{\rm 2}^*$ of the 8-mon (exceeding 30~$\mu$s) substantially surpasses that of conventional X-mons (below 15~$\mu$s), representing a 2–3-fold improvement near the optimal point. At this bias, $T_{\rm 2}^*$ reaches the same order as $T_{\rm 1}$. For example, device $\mathrm{Q_{a3}^*}$ exhibits \(T_{\rm 2}^* \approx 37~\mu\)s while its $T_{\rm 1}$ is about 30~$\mu$s. This significant enhancement directly demonstrates the effectiveness of the gradiometric design in suppressing low-frequency flux noise, which dominates dephasing in this bias regime. The improvement gradually diminishes as the flux bias increases.

The Hahn echo coherence times $T_{\rm 2}^{\rm E}$ for both qubit types are presented in Fig.~\ref{fig3}\bt{c}. In the small flux-bias regime, $T_{\rm 2}^{\rm E}$ for both 8‑mon and X‑mon qubits is on the order of $T_{\rm 1}$, with some devices approaching the 2$T_{\rm 1}$ limit. This comparable performance under echo decoupling indicates that the intrinsic spatial symmetry of the 8‑mon and the dynamical refocusing provided by the echo sequence in the X‑mon are similarly effective in suppressing low‑frequency noise, thereby extending coherence toward the same fundamental limit.

\subsection{Analysis and modeling}

To explain the enhanced coherence of the 8‑mon qubit, we introduce a spatially continuous magnetization model that describes how environmental flux noise couples to the qubit geometry (see Supplementary Information~\ref{theory} for details). 

In this framework, slow surface magnetization fluctuations are treated as a spatially correlated field with a characteristic correlation length \(\xi\). The coupling to the circuit is encoded in a geometry-dependent kernel \(K(\mathbf{r})\). In momentum space, the flux-noise variances for the X‑mon and 8‑mon take the form
\begin{equation*}
\begin{aligned}
  \langle \Phi_{\mathrm{X}}^2\rangle
  &= \int\!\frac{d^2\mathbf k}{(2\pi)^2}\,
     |\tilde K_{\mathrm{X}}(\mathbf k)|^2 \, S_m(\mathbf k;\xi),\\[4pt]
  \langle \Phi_{8}^2\rangle
  &= \int\!\frac{d^2\mathbf k}{(2\pi)^2}\,
     4\sin^2\!\Big(\frac{\mathbf k\cdot\mathbf d}{2}\Big)\,
     |\tilde K_{\mathrm{X}}(\mathbf k)|^2\,
     S_m(\mathbf k;\xi),
\end{aligned}
\end{equation*}
where \(S_m(\mathbf k;\xi)\) is the noise spectrum, \(|\tilde K_{\mathrm{X}}(\mathbf k)|^2\) is the geometric filter of the SQUID of X-mon, and \(\mathbf d\) is the separation between the two sub‑loops of the 8‑mon.

As shown in Figs.~\ref{fig3}\bt{d}–\ref{fig3}\bt{e}, our analysis identifies two distinct noise regimes dictated by the ratio $\xi/d$. In the short‑wavelength limit ($\xi \ll d$), the flux‑noise variances of both the X‑mon and the 8‑mon scale as $\langle\Phi^2\rangle \propto \xi^2$ (Fig.~\ref{fig3}\bt{d}), reflecting locally correlated noise to which the two sub‑loops respond independently~\cite{PhysRevLett.99.187006}. This result is consistent with the approach of geometry-optimized circuits for flux-noise control, where both rely on tailoring the spatial coupling to environmental noise sources~\cite{PhysRevLett.98.267003,PhysRevApplied.13.054079}. In the long‑wavelength limit ($\xi \gtrsim d$), the behaviors diverge markedly. While the X‑mon noise saturates because it couples to spatially uniform field fluctuations~\cite{PhysRevLett.131.070801}, the 8‑mon exhibits a pronounced suppression, $\langle\Phi_8^2\rangle \propto (d/\xi)^2$. This power‑law suppression stems from destructive interference between the two oppositely wound loops, which effectively filters out the dominant low‑frequency components of ambient flux noise. Long-correlation noise has also been reported in Ref.~\cite{PhysRevLett.109.067001}.

These theoretical regimes are directly reflected in the measured Ramsey coherence. The 8‑mon shows a 2–3‑fold enhancement in $T_{\rm 2}^*$ in the small flux-bias regime, where long‑wavelength noise dominates. While the improvement vanishes at larger bias, indicating that both long‑ and short‑wavelength noise sources are present in the experimental environment. In Fig.~\ref{fig3}\bt{f}, we present a numerical simulation of the $T_{\rm 2}^*$ data from Fig.~\ref{fig3}\bt{b}, using a representative noise spectrum that incorporates both correlation limits. The simulation reproduces the measured trend with excellent consistency, confirming that the model captures the essential spatial filtering provided by the gradiometric geometry.

%%%%%%%%%%%%%%%%%%%%%% FIG. 4 %%%%%%%%%%%%%%%%%%%%%%%%
\begin{figure}[ht]
\includegraphics[width=\linewidth]{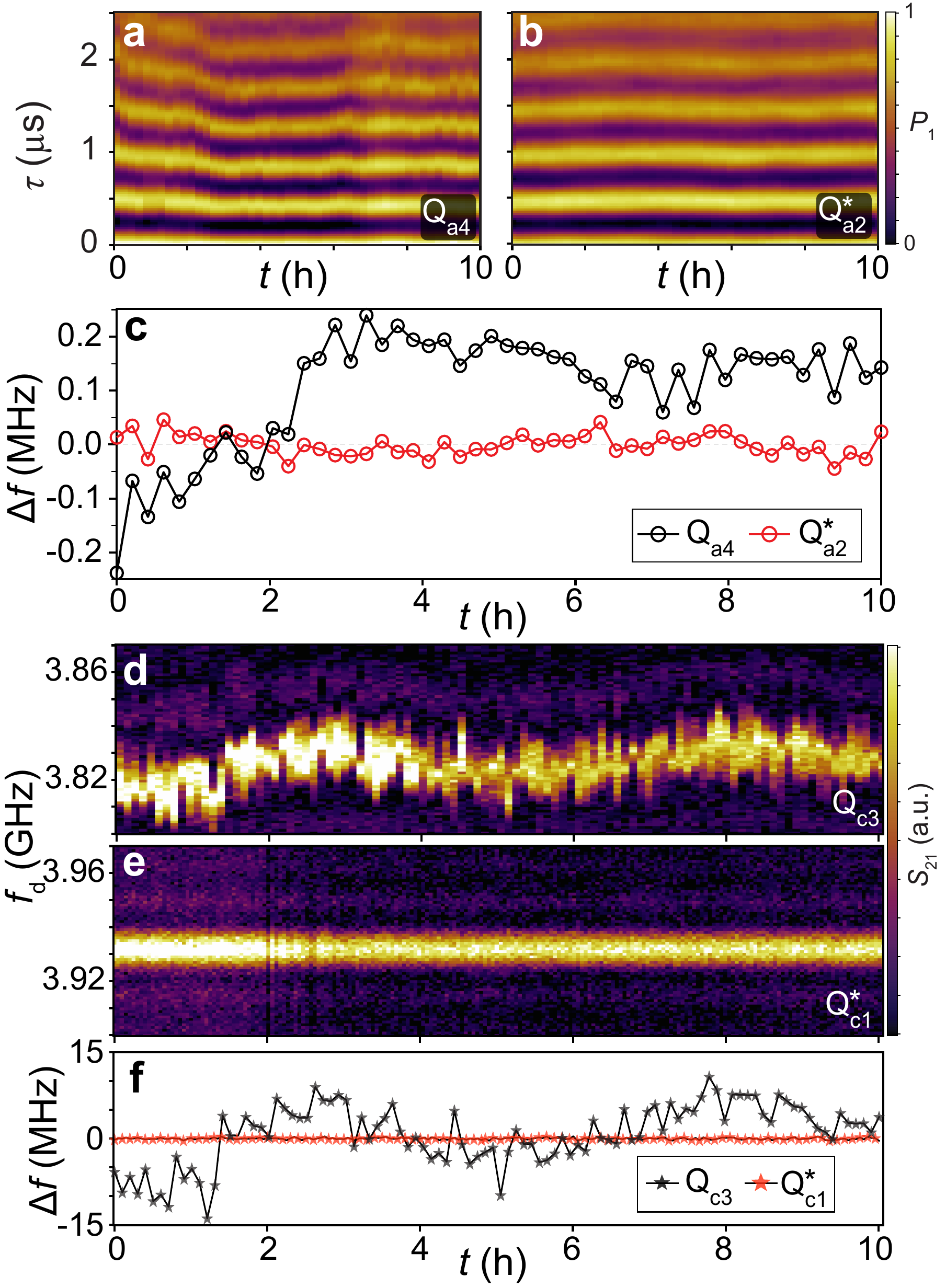}
\caption{\bt{Long-term frequency stability of 8-mon and X-mon qubits.} \bt{a-b}, Representative Ramsey fringes measured over 10 hours for a reference X-mon (\bt{a}) and an 8-mon (\bt{b}) qubit, both biased at 0.1 $\Phi_0$. \bt{c}, Qubit frequency drift extracted from the Ramsey measurements in \bt{a} and \bt{b}. The 8-mon qubit exhibits superior stability with a drift $<0.1$ MHz, markedly lower than the $\approx 0.4$ MHz drift of the X-mon. \bt{d-e}, Direct spectroscopic monitoring of the qubit transition frequency $f_{\rm 01}$ over 10 hours for an X-mon (\bt{d}) and an 8-mon (\bt{e}) in the absence of a cryogenic magnetic shield. \bt{f}, Frequency drift extracted from Gaussian fits to the spectroscopic data in \bt{d} and \bt{e}. Without shielding, the X-mon experiences a drastic frequency drift exceeding 20 MHz, whereas the 8-mon maintains remarkable stability with no observable drift.}
\label{fig4}
\end{figure}
%%%%%%%%%%%%%%%%%%%%%%%%%%%%%%%%%%%%%%%%%%%%%%%%%%%

\subsection{Long-term frequency stability}
The long-term frequency stability of the 8‑mon was characterized alongside reference X‑mons, highlighting a decisive advantage of the gradiometric architecture. Under a fixed flux bias of 0.1~$\Phi_0$ and with cryogenic magnetic shielding, the 8‑mon exhibited a frequency drift of less than 0.1~MHz over 10~hours (Figs.~\ref{fig4}\bt{a}–\ref{fig4}\bt{c}). In contrast, a conventional X‑mon under the same conditions drifted by approximately 0.4~MHz. 

Without magnetic shielding, the difference becomes even more pronounced (Figs.~\ref{fig4}\bt{d}-\ref{fig4}\bt{f}). The X-mon exhibited severe frequency instability exceeding 20~MHz, whereas the 8-mon maintained exceptional stability with no observable drift. These results demonstrate that the doubly connected gradiometric topology inherently suppresses ambient low-frequency magnetic field fluctuations, thereby passively eliminating the long-term frequency drift that plagues standard tunable transmons. Unlike active, feedback-based frequency locking techniques~\cite{Vepsalainen2022, berritta2025efficient}, our approach eliminates the need for additional control overhead while providing intrinsic immunity to environmental fluctuations.
\vspace{0.3cm}
\section*{Discussion}

Our work demonstrates that the doubly connected gradiometric transmon significantly improves coherence while preserving full frequency tunability. The nano-airbridge interloop connection plays a key role in maintaining high energy-relaxation times $T_{\rm 1}$ comparable to conventional X-mons, by eliminating the dielectric loss associated with amorphous SiO\(_2\) layers. Under small flux bias, the 8-mon exhibits a nearly threefold enhancement in Ramsey coherence time $T_{\rm 2}^*$ over reference X-mon qubits, reaching the same order as $T_{\rm 1}$ without employing dynamical decoupling sequences. The long-term frequency stability is equally improved, with a drift below 0.1 MHz over 10 hours, a fourfold reduction compared to conventional devices. These experimental observations are quantitatively captured by a spatially correlated flux-noise model, which reveals the coexistence of short- and long-correlation-length magnetic noise in the superconducting chip environment. The geometric suppression of low-frequency flux fluctuations does not compromise electrical tunability, nor does it require additional control complexity, offering a practical route toward more robust and scalable superconducting quantum processors.

Several promising directions can further advance the performance of gradient-based transmons. Improving the gradiometric symmetry through narrower superconducting arms would homogenize current distribution and enhance common-mode rejection. Extending the cancellation principle into the third dimension, via a three-dimensional gradiometric topology, could mitigate sensitivity to out-of-plane magnetic flux variations. Integrating the 8-mon design with low-loss superconducting materials such as tantalum-based films may substantially increase $T_{\rm 1}$ and further push the limit of $T_{\rm 2}^*$ performance~\cite{Place2021, Wang2022, Li2023, Bland2025}. Additionally, exploring hybrid architectures that combine the 8-mon with tunable couplers or multi-qubit lattices would test its scalability and compatibility with existing fabrication processes. These efforts will help clarify the impact of material-limited relaxation on dephasing and solidify the role of geometric noise filtering in next-generation quantum processors.

\section*{Data availability}
Relevant data supporting the key findings of this study are available within the article and the Supplementary Information file. All raw data generated during the current study are available from the corresponding authors upon reasonable request.

\newpage
\bibliography{bibfile}
\bibliographystyle{naturemag}

\section*{Acknowledgments}
We are grateful to Yang Yang, Xiao-Feng Yi, Jian-Dong Ouyang, Peng Luo, Kang-Ding Zhao, and Wei-Chen Wang for their technical support in device fabrication and qubit measurements. This work was funded by the National Key R\&D Program of China (Grant No. 2024YFB4504000), and the Aid Program for Science and Technology Innovative Research Team in Higher Educational Institutions of Hunan Province. Device fabrication was partially carried out at the Synergetic Extreme Condition User Facility (SECUF).

\section*{Author Contributions} 
M.-T.D. conceived the central concepts and coordinated the experimental work. Device fabrication was carried out by J.-B.F. and B.R., while J.-B.F. and M.-T.D. performed the measurements. D.-W.W. and Y.-X.L. conducted the theoretical modeling and numerical simulation simulations. S.H., D.-D.L., J.-B.F., and Z.-H.Y. developed the software under the supervision of X.F. All authors participated in interpreting the results and in the preparation of the manuscript.

\clearpage

\setcounter{section}{0}
\renewcommand{\thesection}{S\arabic{section}}

\vspace{5em}
\onecolumngrid
\begin{center}
    \textbf{\large Supplementary Information}
\end{center}
\vspace{0.5em}

\maketitle
\setcounter{figure}{0}
\renewcommand{\thefigure}{S\arabic{figure}}

% \captionsetup[figure]{name=Fig. S}

% %%%%%%%%%%%%%%%%%%%%%% FIG. S1 %%%%%%%%%%%%%%%%%%%%%%%%
% \begin{figure}[t]
% \includegraphics[width=0.75\linewidth]{FigS1-3.pdf}
% \caption{\bt{Experimental conditions.} \bt{a}, Chip-A conditions,including magnetic shielding box}
% \label{figS1}
% \end{figure}
% %%%%%%%%%%%%%%%%%%%%%%%%%%%%%%%%%%%%%%%%%%%%%%%%%%%

\section{Device and Measurement Setup}
\label{Setup}
\begin{figure}[h]
\includegraphics[width=10cm]{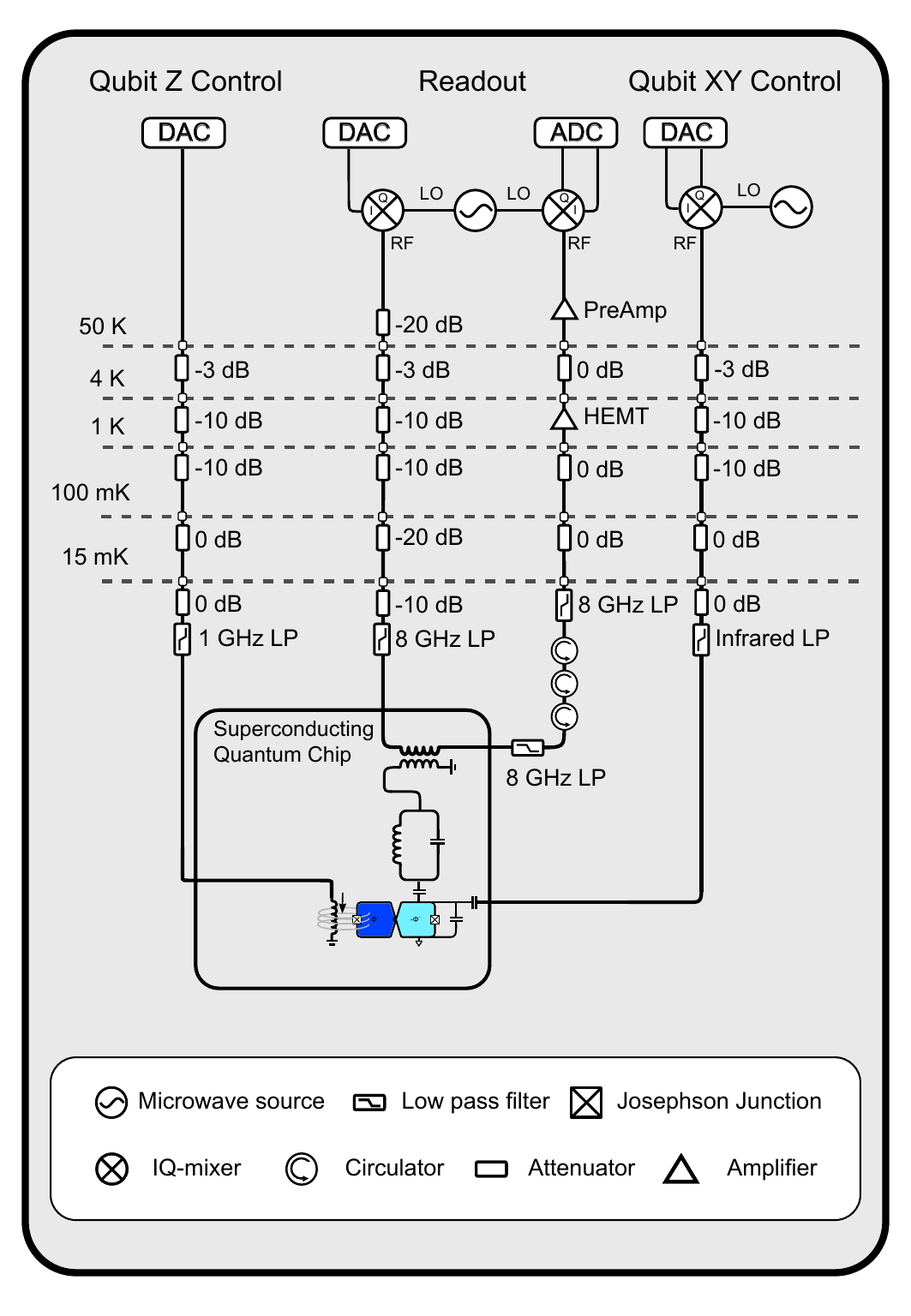}
\caption{\bt{Wiring schematic of measurement setup.} }
\label{figS1}
\end{figure}

Figure~\ref{figS1} illustrates the experimental setup, including the DC and microwave components within the dilution refrigerator system used in all measurements. Qubit XY control and readout pulses are generated by an arbitrary waveform generator. The in-phase and quadrature pulses are up-converted to the qubit and resonator frequencies using two radio-frequency sources, each coupled to an external IQ mixer for single-sideband modulation. The signals are attenuated at successive cooling stages to suppress thermal noise from higher-temperature stages, and are then delivered to the superconducting quantum chip mounted on the mixing-chamber plate of the dilution refrigerator, which operates at a base temperature of approximately 15~\text{mK}. Notably, neither Josephson parametric amplifiers (JPA) nor traveling-wave parametric amplifiers (TWPA) were used in any of the three experiments presented in the main article.

In this experiment, three chips (labeled chip‑\textit{a}, chip‑\textit{b}, and chip‑\textit{c}) were measured. Chip‑\textit{a} and chip‑\textit{b} were each equipped with a permalloy magnetic shield to suppress external magnetic field fluctuations, whereas chip‑\textit{c} was operated without shielding to directly assess its impact on qubit frequency stability.

%%%%%%%%%%%%%%%%%%%%%% FIG. S1 %%%%%%%%%%%%%%%%%%%%%%%%
%%%%%%%%%%%%%%%%%%%%%%%%%%%%%%%%%%%%%%%%%%%%%%%%%%%

\section{Qubit Spectrum and Readout}
\label{Readout}
\begin{figure}[t]
\includegraphics[width=14cm]{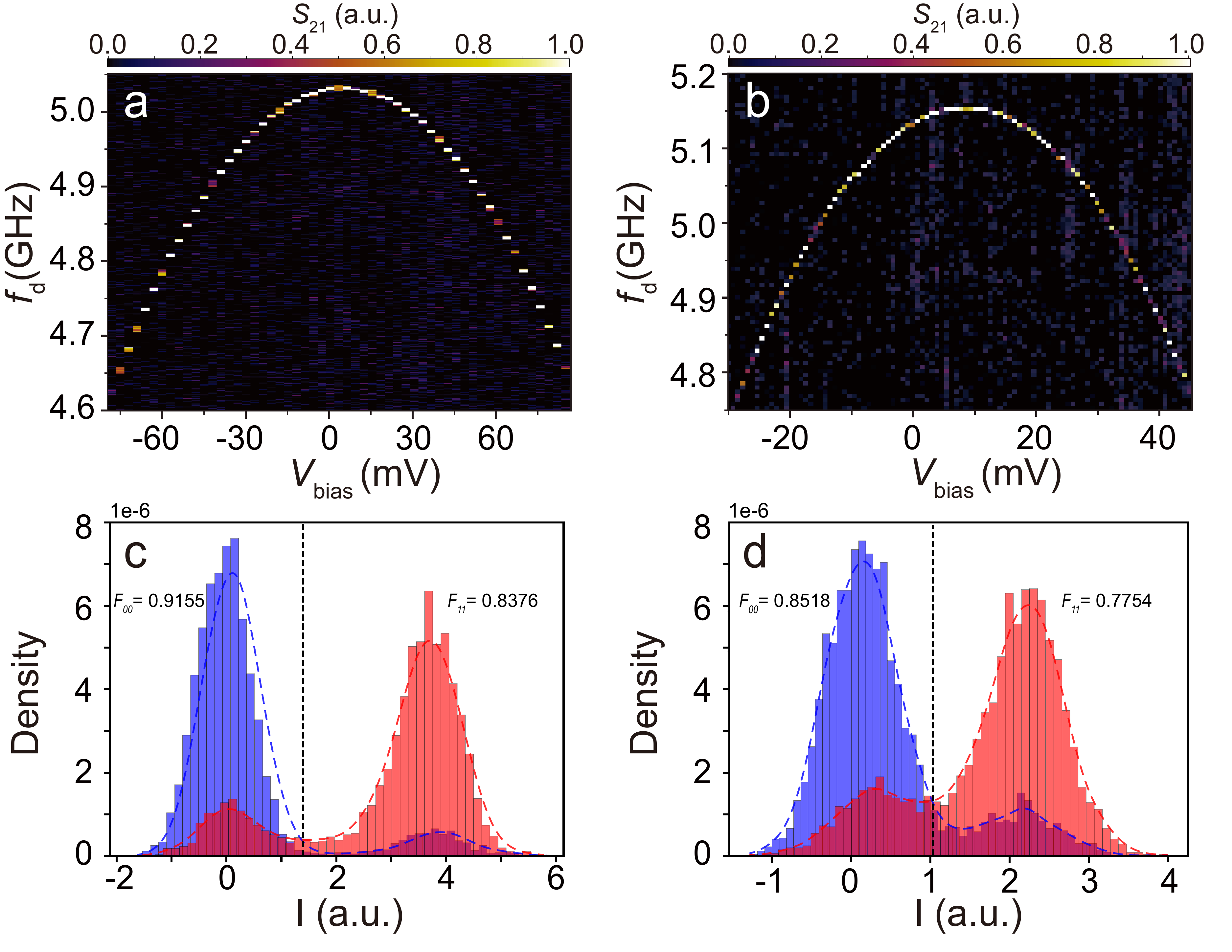}
\caption{\bt{Energy spectrum of 8-mon and X-mon qubits.} \bt{a}, 8-mon qubit's spectrum with $f_{01}$ = 5.028 GHz at optimization Point. \bt{b}, X-mon qubit's spectrum with $f_{01}$ = 5.150 GHz at optimization Point. \bt{c}, 8-mon qubit's  readout fidelity experiment. \bt{d}, X-mon qubit's readout fidelity experiment.}
\label{figS2}
\end{figure}

Since the 8‑mon and X‑mon share identical Hamiltonians, their energy spectra and frequency‑tuning ranges are essentially equivalent, as shown in Figs.~\ref{figS2}\bt{a}-\ref{figS2}\bt{b}. Moreover, both qubits exhibit highly similar responses to the applied bias voltage (current). These results confirm that the gradiometric design preserves the full electrical tunability inherent to standard transmon qubits. 

In terms of readout fidelity, both the 8-mon and the X-mon employ the dispersive readout scheme. As shown in Figs.~\ref{figS2}\bt{c} and \ref{figS2}\bt{d}, the 8-mon exhibits comparable performance to the X-mon. Without the use of a JPA or TWPA, and prior to any readout correction or optimization, the measured readout fidelity for both devices is approximately 80\%--90\%.

%%%%%%%%%%%%%%%%%%%%%% FIG. S2 %%%%%%%%%%%%%%%%%%%%%%%%
%%%%%%%%%%%%%%%%%%%%%%%%%%%%%%%%%%%%%%%%%%%%%%%%%%%

% \renewcommand{\thefigure}{S\arabic{figure}}
% \setcounter{figure}{0} 

%%%%%%%%%%%%%%%%%%%%%% FIG. S4 %%%%%%%%%%%%%%%%%%%%%%%%
\begin{figure*}[t]
\includegraphics[width=\linewidth]{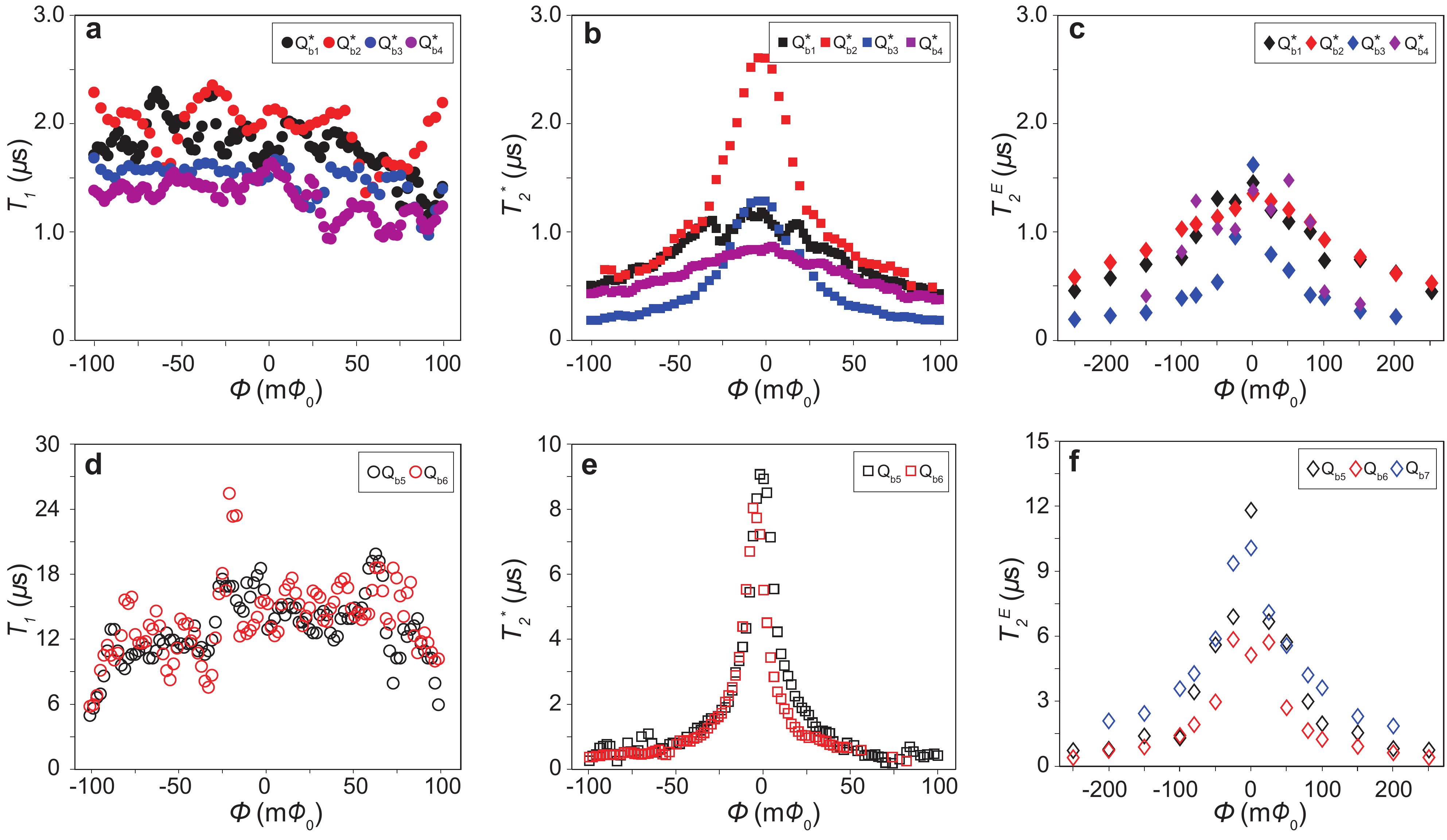}
\caption{\bt{Coherence time characterization for qubits on chip-\emph{b}.} \bt{a}, Energy relaxation time $T_{1}$ for four 8-mon qubits. \bt{b}, Ramsey coherence time $T_{2}^{*}$ of the 8-mon as a function of the applied flux bias. \bt{c}, Echo decoherence time $T_{2}^{\mathrm{E}}$ of the 8-mon as a function of the applied flux bias. \bt{d-f}, Corresponding measurements of $T_{1}$, $T_{2}^{*}$, and $T_{2}^{\mathrm{E}}$ for X-mon qubits on the same chip. Both $T_{1}$ and $T_{2}$ of the 8-mon are limited by dielectric loss.}
\label{figS3}
\end{figure*}
%%%%%%%%%%%%%%%%%%%%%%%%%%%%%%%%%%%%%%%%%%%%%%%%%%%

\section{Coherence time characterization for qubits on chip-\textit{b}}
\label{S4}
 Chip‑\textit{b} is largely identical to chip‑\textit{a} in overall layout and qubit geometry, except for the cross‑bridge design. Here, a SiO$_2$ layer serves both as the bridge's supporting structure and as an insulating dielectric. 

 In Figure~\ref{figS3}, we compared the coherence times of 8‑mon and X‑mon qubits on chip‑\textit{b}. The $T_1$ of the 8‑mon is about 1--2~$\mu$s, much lower than that of X-mon. This can be attributed to the loss introduced by the SiO$_2$ layer in 8-mon structure. For the short lifetime, the $T_2^*$ and $T_2^E$ of 8-mon on chip‑\textit{b} are under 3 $\mu$s correspondingly(Figs.~\ref{figS3}\bt{b}-\ref{figS3}\bt{c}). In contrast, the $T_1$ of the X‑mon on chip‑\textit{b}—which does not incorporate this dielectric—remains similar to that of the X‑mons on chip‑\textit{a}. 

Comparing chip-\textit{a} and chip-\textit{b}, we find that avoiding a dielectric layer is critical for the 8-mon performance.

\section{Details of theoretical model}
\label{theory}
\subsection{ General framework}

We consider the response of a superconducting loop geometry, such as an X-mon loop or an 8-mon loop,
to magnetic flux noise generated by surface magnetization fluctuations.
In general, magnetic noise is characterized by both spatial and temporal correlations ~\cite{PhysRevLett.131.070801}.
In this work, however, we focus on the spatial structure of the noise and its interplay with circuit geometry,
while treating the temporal fluctuations implicitly.

Specifically, we consider the quasi-static or low-frequency regime relevant for Ramsey dephasing experiments,
where the characteristic fluctuation time of the magnetization noise is long compared to the qubit evolution time.
In this regime, the dominant contribution to dephasing is governed by the time-integrated flux variance,
and the effect of temporal correlations can be absorbed into an overall noise amplitude.
This allows us to isolate the geometry-dependent contribution to flux noise
without specifying the detailed microscopic origin of the temporal noise spectrum.
Under these assumptions, the magnetic flux threading the loop can be expressed in the general form~\cite{PhysRevB.92.054502,PhysRevLett.100.227005}
\begin{equation}
  \Phi
  =
  \int d^2\mathbf r\; K(\mathbf r)\, m_z(\mathbf r),
  \label{eq:flux_definition}
\end{equation}
where $m_z(\mathbf r)$ denotes the out-of-plane component of the surface magnetization field,
and $K(\mathbf r)$ is the geometry-dependent sensitivity kernel,
fully determined by the loop layout and encoding the contribution of a unit magnetization at position $\mathbf r$ to the total magnetic flux.
The second-order spatial correlation function of the magnetization noise is defined as
\begin{equation}
  C_m(\mathbf r-\mathbf r';\xi)
  =
  \big\langle
    m_z(\mathbf r)\, m_z(\mathbf r')
  \big\rangle,
\end{equation}
where the angular brackets denote an ensemble average over slowly fluctuating magnetization configurations,with temporal correlations implicitly integrated out.
The parameter $\xi$ denotes the characteristic spatial correlation length,
which can be physically interpreted as the typical size of magnetic domains or correlated spin clusters.

The corresponding noise spectrum in momentum space is given by the two-dimensional Fourier transform
\begin{equation}
  S_m(\mathbf k;\xi)
  =
  \int d^2\mathbf r\;
  C_m(\mathbf r;\xi)\,
  e^{-i\mathbf k\cdot\mathbf r}.
\end{equation}
Similarly, we define the Fourier transform of the geometry kernel as
$\tilde K(\mathbf k)  = \int d^2\mathbf r\; K(\mathbf r)\,e^{-i\mathbf k\cdot\mathbf r}$.
Then the magnetic flux variance $\langle \Phi^2\rangle$,
which quantifies the effective strength of low-frequency flux noise relevant for dephasing,
is most transparently evaluated in momentum space.
Using the convolution theorem and standard Fourier identities, one obtains
\begin{equation}
  \langle \Phi^2\rangle
  =
  \int\frac{d^2\mathbf k}{(2\pi)^2}\;
  \big| \tilde K(\mathbf k)\big|^2\,
  S_m(\mathbf k;\xi).
  \label{eq:Phi2_k_space_master}
\end{equation}

Equation~\eqref{eq:Phi2_k_space_master} constitutes a central result of this work.
It shows that the loop geometry acts as a spatial filter in momentum space,
weighting the magnetization noise spectrum $S_m(\mathbf k)$
by the geometry-dependent filter function $\big| \tilde K(\mathbf k)\big|^2$.
Importantly, this formulation isolates the role of circuit geometry in suppressing
or enhancing low-frequency flux noise,
independent of the detailed temporal structure of the underlying noise processes.

\subsection{Geometry kernels}
 \begin{figure}[h]
	\begin{center}
		\includegraphics[width=10cm]{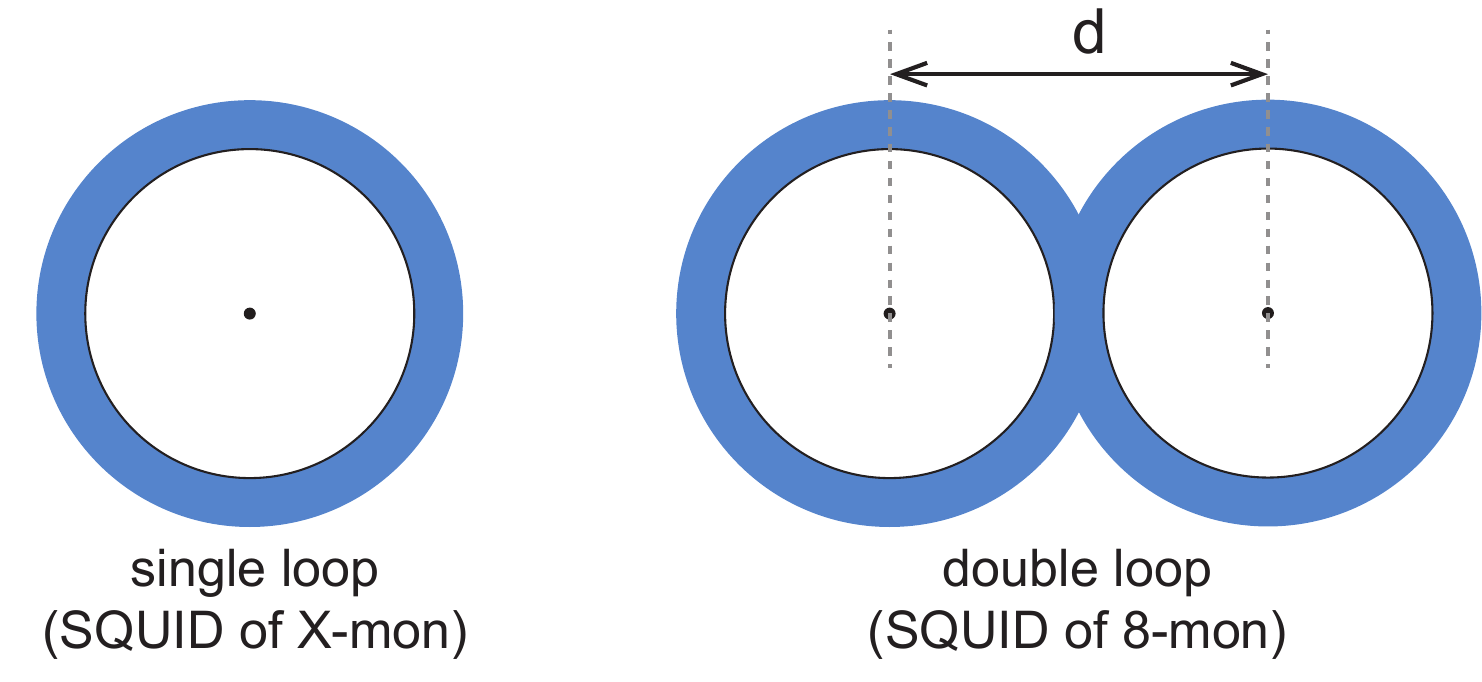}
\caption{
Schematic diagram of the loop geometry considered in the simulation.
Left: A  single circular ring in an X-mon  circuit  configuration.
Right: Two identical rings with opposite current directions in 8-mon  circuit  configuration. The centers of the two sub-rings are separated by a distance $d$.
}
\label{fig:model}
	\end{center}
\end{figure}
To facilitate analysis while preserving the basic geometric features of the actual devices, we approximate the SQUID layout of the two experimental circuits as an idealized ring geometry. Similar geometric approximations have been adopted in previous study~\cite{PhysRevLett.100.227006}.
The X-mon circuit uses a single circular ring, while the 8-mon circuit uses two identical rings with opposite circulation directions, as shown in Fig.~\ref{fig:model}.
We denote by $K_{\rm ring}(\mathbf r)$ the geometry kernel of a single ring centered at the origin, and by $\tilde K_{\rm ring}(\mathbf k)$ its Fourier transform.
A spatial translation by a vector $\mathbf R$ corresponds in momentum space to a phase factor,
$K_{\rm ring}(\mathbf r-\mathbf R)\to \tilde K_{\rm ring}(\mathbf k)e^{-i\mathbf k\cdot\mathbf R}$.

For the X-mon circuit, the geometric kernel coincides with that of a single ring as
$K_{\rm X}(\mathbf r)= K_{\rm ring}(\mathbf r)$ and  $ \tilde K_{\rm X}(\mathbf k)= \tilde K_{\rm ring}(\mathbf k)$.
The 8-mon circuit  consists of two identical rings centered at
$\mathbf R_L=-\mathbf d/2$ and $\mathbf R_R=\mathbf d/2$, carrying opposite circulating currents.
Its geometric kernel is therefore
$K_{8}(\mathbf r)= K_{\rm ring}(\mathbf r-\mathbf R_L)- K_{\rm ring}(\mathbf r-\mathbf R_R)$.
Taking the Fourier transform yields $\tilde K_{8}(\mathbf k)=2i\,\tilde K_{\rm ring}(\mathbf k)\sin\!\Big(\mathbf k\cdot\mathbf d/2\Big)$, where $\mathbf d=\mathbf R_R-\mathbf R_L$ is the center-to-center separation between the two rings.
Consequently, the geometry-dependent weights between the X-mon and 8-mon circuits satisfy
\begin{equation}
  \big|\tilde K_8(\mathbf k)\big|^2
  =
  4\,\big|\tilde K_{\rm X}(\mathbf k)\big|^2
  \sin^2\!\Big(\frac{\mathbf k\cdot\mathbf d}{2}\Big),
  \label{eq:K8-vs-Ks}
\end{equation}
which forms the basis for our analysis of geometry-induced noise suppression.
Substituting Eq.~\eqref{eq:K8-vs-Ks} into Eq.~\eqref{eq:Phi2_k_space_master}, the flux-noise variances for the X-mon and 8-mon circuits are
\begin{subequations}
\begin{align}
  \langle \Phi_{\rm s}^2\rangle
  &= \int\!\frac{d^2\mathbf k}{(2\pi)^2}\,
     W(\mathbf k)\, S_m(\mathbf k;\xi),\\
  \langle \Phi_{8}^2\rangle
  &= \int\!\frac{d^2\mathbf k}{(2\pi)^2}\,
     4\,W(\mathbf k)
     \sin^2\!\Big(\frac{\mathbf k\cdot\mathbf d}{2}\Big)
     S_m(\mathbf k;\xi),
\end{align}
\end{subequations}
where \(W(\mathbf k)\equiv|\tilde K_{\rm X}(\mathbf k)|^2\).
Next we define the geometry-induced suppression factor
\begin{equation}
  \mathrm{S}(\xi)
  \equiv
  \frac{\langle \Phi_{\rm X}^2\rangle}
       {\langle \Phi_{8}^2\rangle}.
\end{equation}
And  introducing the weighted average
\begin{equation}
  \langle f(\mathbf k)\rangle_\xi
  =
  \frac{\int \frac{d^2\mathbf k}{(2\pi)^2}\,
        W(\mathbf k)S_m(\mathbf k;\xi)f(\mathbf k)}
       {\int \frac{d^2\mathbf k}{(2\pi)^2}\,
        W(\mathbf k)S_m(\mathbf k;\xi)},
\end{equation}
Then the suppression factor can be rewritten as 
\begin{equation}
  \mathrm{S}(\xi)
  =
  \frac{1}{
    4\langle \sin^2(\mathbf k\cdot\mathbf d/2)\rangle_\xi}.
  \label{eq:Supp-average}
\end{equation}
Therefore, the behavior of the suppression factor is thus governed by the ratio of the dominant noise wavelength to the inter-ring distance $d$. This gives rise to two distinct physical regimes: the short-wavelength-dominated and long-wavelength-dominated regimes as below.
\begin{itemize}
\item [(1)]\textbf{Short-wavelength--dominated regime.}

In this regime, the spectral weight of the magnetization noise $S_m(\mathbf k)$ is concentrated at wave vectors $k\gg 1/d$, corresponding to fluctuations whose effective magnetization patterns vary on length scales much shorter than the inter-ring distance.
This class includes uncorrelated (independent-spin) noise, for which
$C_m(\mathbf r)=A_0\delta(\mathbf r)$ and $S_m(\mathbf k)=S_0$, as well as correlated noise with substantial high-$k$ components.
The geometric interference factor
$\sin^2(\mathbf k\cdot\mathbf d/2)$
oscillates rapidly as a function of the orientation of $\mathbf{k}$ relative to the separation vector $\mathbf{d}$. For a noise ensemble with an approximately isotropic distribution, the weighted average over these rapid oscillations simplifies to $\langle\sin^2(\mathbf{k}\cdot\mathbf{d}/2)\rangle_\xi\approx1/2$.
Substituting this result into Eq. \eqref{eq:Supp-average}, we find that the suppression factor $S$ is of order unity:
\begin{equation}
  \mathrm S(\xi)
  \simeq
  \frac{1}{4\langle \sin^2(\mathbf k\cdot\mathbf d/2)\rangle_\xi}
  \approx
  \frac{1}{2}.
\end{equation}
Therefore, under conditions dominated by short-wavelength magnetization fluctuations, the phase correlation between the two sub-rings practically disappears. In this limiting case, the 8-mon circuit offers no advantage in suppressing flux noise, and its performance is comparable to that of  the X-mon circuit.

\item [(2)]\textbf{Long--wavelength–dominated regime.}

In the opposite limit, the magnetization noise is dominated by wavelengths much longer than the sub-rings separation, such that the spectral weight of $S_m(\mathbf k)$ is concentrated at $k\ll 1/d$.
This situation arises when the magnetization varies smoothly across the device.
Then the interference factor can be expanded as  $\sin^2\!(\mathbf k\cdot\mathbf d/2) \simeq(\mathbf k\cdot\mathbf d/2)^2$.
For an approximately isotropic weighted integral dominated by wave vectors
$|\mathbf k|\sim k_{\rm dom}\ll 1/d$, this leads to
$\mathrm S(\xi)\sim[ d^2 k_{\rm dom}^2]^{-1}$.
For noise characterized by a single spatial correlation length $\xi$, such that
$k_{\rm dom}\sim 1/\xi$, one obtains the asymptotic scaling
\begin{equation}
  \mathrm S(\xi)
  \propto
  \Big(\frac{\xi}{d}\Big)^2.
  \label{eq:Supp-long}
\end{equation}
Therefore, in the long-wavelength regime of magnetization noise, the 8-mon circuit provides a more effective noise suppression than the X-mon circuit.
\end{itemize}

Our analysis shows that the flux-noise suppression ratio $  \mathrm S(\xi)$
of the 8-mon circuit relative to the X-mon circuit is governed by a simple spatial criterion: the comparison between the characteristic wavelength of magnetization noise and the sub-ring separation $d$. For short-wavelength fluctuations, the two sub-rings respond almost independently, leading only to a weak reduction of noise. In contrast, when the magnetization varies smoothly over length scales larger than $d$, the opposite circulating currents in the two sub-rings give rise to strong destructive interference, resulting in a power-law enhancement of the suppression ratio.
This scaling behavior is unique to the 8-mon circuit geometry and does not rely on any specific microscopic noise model. It reflects the role of the 8-mon circuit as a spatial filter that selectively suppresses the long-wavelength components of magnetization noise.   Conceptually, our approach can be viewed as a form of spatial noise filtering, analogous to temporal noise spectroscopy using qubit probes. While noise spectroscopy filters environmental fluctuations in frequency space through control sequences, the figure-8 geometry of 8-mon filters noise in momentum space through circuit geometry. Both approaches rely on the same underlying principle of probing noise correlations via controlled filtering~\cite{PhysRevLett.131.070801}.

\subsection{ Numerical simulation}
In the numerical calculations, we adopt a representative magnetization noise spectrum with a finite spatial correlation length,
\begin{equation}
S_m(\mathbf k;\xi)\propto \xi^2\, e^{-k^2\xi^2},
\end{equation}
which is peaked around wave vectors $|\mathbf k|\sim 1/\xi$ and provides a smooth interpolation between short  and long wavelength–dominated noise regimes.
The prefactor $\xi^2$ reflects the scaling of the effective magnetization strength of a correlated region with its area in two dimensions.

In the short-wavelength regime $kd\gg1$,  namely $\xi\ll d$, the magnetization noise spectrum is effectively flat over the momentum range relevant to the geometric kernel. The flux-noise variance of X-mon circuit then reads
\begin{equation}
\langle \Phi_{\rm X}^2 \rangle
=
\int \frac{d^2\mathbf{k}}{(2\pi)^2}\,
\big|\tilde K_{\rm X}(\mathbf{k})\big|^2\,
S_m(\mathbf{k})
\propto
\xi^2
\int \frac{d^2\mathbf{k}}{(2\pi)^2}\,
\big|\tilde K_{\rm X}(\mathbf{k})\big|^2 ,
\end{equation}
where the dependence on $\xi$ enters only through an overall prefactor.
For the 8-mon circuit, using
$\big|\tilde K_{\rm 8}(\mathbf{k})\big|^2
=4\big|\tilde K_{\rm X}(\mathbf{k})\big|^2
\sin^2(\mathbf{k}\cdot\mathbf d/2)$,
one similarly finds
\begin{equation}
\langle \Phi_{\rm 8}^2 \rangle
\propto
2\xi^2
\int \frac{d^2\mathbf{k}}{(2\pi)^2}\,
\big|\tilde K_{\rm X}(\mathbf{k})\big|^2 .
\end{equation}
Therefore, in the short-wavelength regime, the flux-noise  of both circuits scale as $\langle \Phi^2\rangle\propto\xi^2$ in agreement with the linear increase observed at small $\xi$ in Fig.~\ref{fig3}\bt{d}. 
Physically, the short-wavelength regime corresponds to the independent noise limit, where magnetization fluctuations are essentially uncorrelated on the circuit scale.
Under this limit, the flux-noise variance is primarily governed by the geometry-dependent kernel function
and scales with the effective loop circumference~\cite{PhysRevLett.99.187006,PhysRevLett.100.227005}.
Consequently, the flux noise amplitude in an 8-mon circuit is twice that of a X-mon circuit, consistent with the 8-mon geometry possessing twice the circumference.

In the long-wavelength regime $\xi\gtrsim d$, the noise spectrum becomes sharply
concentrated around $k\simeq 0$.
Then for the X-mon circuit,  the geometry kernel becomes
$\big|\tilde K_{\rm X}(\mathbf{k})\big|^2
\xrightarrow{k\to 0}
\big|\tilde K_{\rm X}(0)\big|^2$
indicating that uniform (common-mode) magnetization fluctuations couple
directly to the ring.
Consequently,
\begin{equation}
\langle \Phi_{\rm X}^2 \rangle
\xrightarrow{\xi \gg d}
\text{const},
\label{eq:single_plateau}
\end{equation}
leading to a saturation (plateau) of the flux noise at large $\xi$. Physically, once the magnetization is correlated over length scales comparable to or exceeding the ring size, further increasing the correlation length does not increase the net magnetic flux threading the loop.
In contrast, for the 8-mon circuit geometry the corresponding kernel satisfies
\begin{equation}
\big|\tilde K_{8}(\mathbf{k})\big|^2
\xrightarrow{k\to 0}
\big|\tilde K_{\rm X}(0)\big|^2\,
(\mathbf k\cdot\mathbf d)^2
\propto k^2\big|\tilde K_{\rm X}(0)\big|^2.
\label{eq:K8_k2_smallk}
\end{equation}
This explicitly shows that uniform magnetization fluctuations do not couple to the 8-mon circuit, as the common-mode contribution is canceled by the opposite circulation of the two rings.
As a result, in the long-wavelength limit the flux-noise  of the
8-mon circuit scales as
\begin{equation}
\langle \Phi_{8}^2 \rangle
\propto
\int d^2\mathbf{k}\, k^2\, S_m(\mathbf{k})
\sim
\Big(\frac{d}{\xi}\Big)^2,
\quad (\xi\gg d),
\label{eq:fig8_long}
\end{equation}
and therefore decreases with increasing correlation length in agreement with the linear increase observed at small $\xi$ in Fig.~\ref{fig3}\bt{d}.

\subsection{Extraction of Ramsey Dephasing under  Flux Noise}
\label{app:ramsey_flux_noise}
The qubit transition frequency depends on the applied magnetic flux, $\omega=\omega(\Phi)$.
Flux fluctuations $\delta\Phi$ induce frequency fluctuations $\delta\omega(\Phi)=\omega(\Phi+\delta\Phi)-\omega(\Phi)$.
Expanding around a given bias point to second order yields
\begin{equation}
\delta\omega(\Phi)
\simeq
D_1(\Phi)\,\delta\Phi
+
\frac{1}{2}D_2(\Phi)\,\delta\Phi^2,
\end{equation}
with  $D_1(\Phi)=\partial_\Phi\omega$ and $ D_2(\Phi)=\partial_\Phi^2\omega$.
For symmetric SQUID-based devices, the flux sweet spot satisfies
$D_1(0)\approx0$, such that the quadratic term dominates in its vicinity.  We assume quasi-static flux noise, i.e., $\delta\Phi$ is constant during a single Ramsey evolution but varies between repetitions, and is Gaussian distributed
with $\delta\Phi \sim \mathcal N(0,\sigma_\Phi^2)$,
where $\sigma_\Phi$ denotes the rms flux noise amplitude. In the quasi-static limit, the accumulated phase is
$\varphi(t)=\delta\omega(\Phi)\,t$,
and the flux-noise-induced Ramsey coherence factor is given by $W_{\rm flux}(t)=\big\langle e^{i\varphi(t)} \big\rangle$.
 \begin{figure}[h]
	\begin{center}
		\includegraphics[width=14cm]{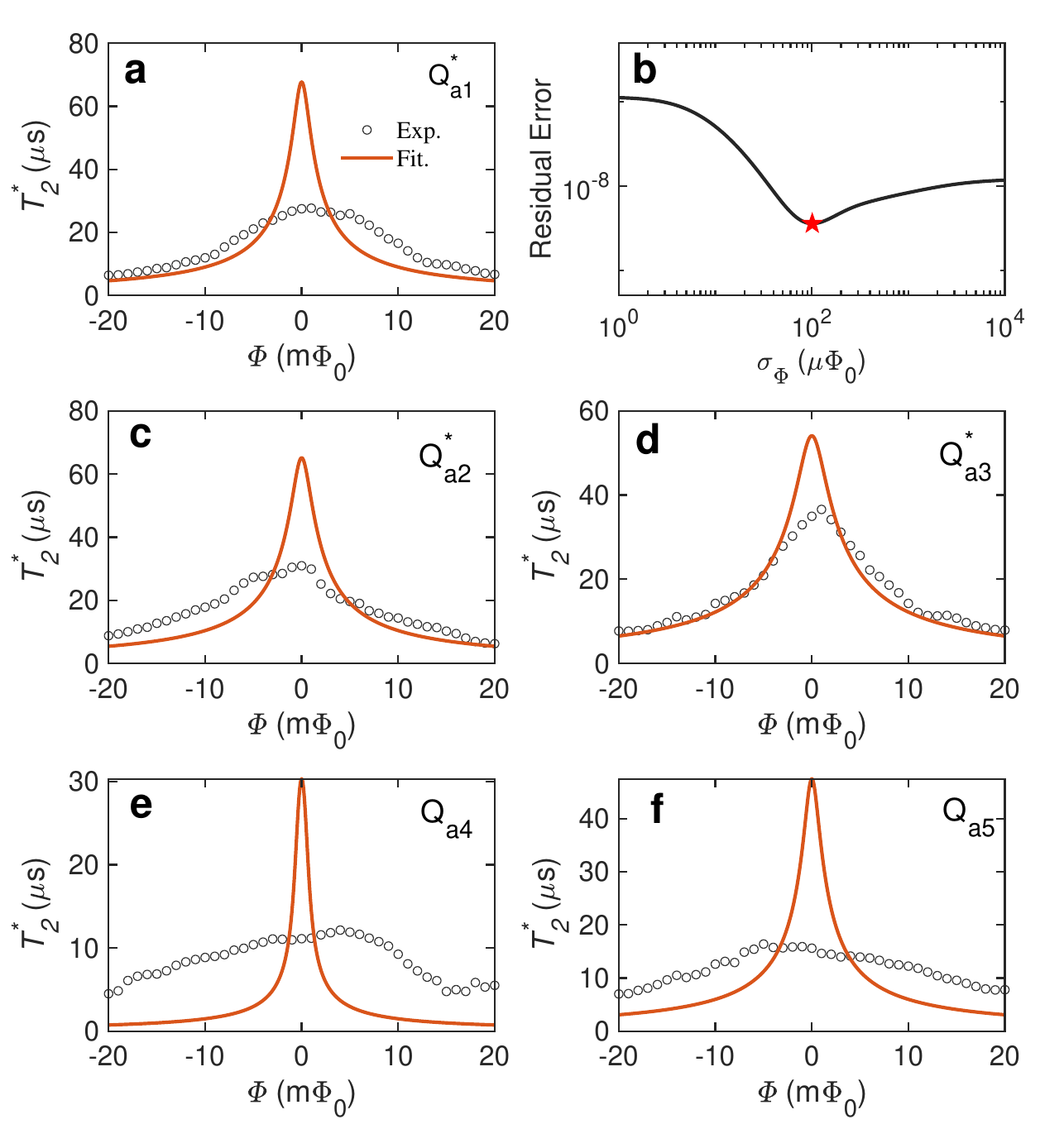}
\caption{\bt{Bias-dependent flux noise analysis of Ramsey decoherence in multiple devices.} 
Symbols show the measured Ramsey dephasing time $T_2^*$ as a function of flux bias, while solid lines represent fits based on quasi-static flux noise including both first- and second-order flux contribution.
Panel \bt{a} shows the flux-noise--only fit for device  $\rm Q_{\rm a1}^*$, and  the  panel \bt{b} displays the corresponding error landscape used to extract the effective flux-noise amplitude $\sigma_\Phi$.
The  panels  \bt{c}-\bt{f} present the experimental data and independent flux-noise--only fits for the other devices. Devices  $\rm Q_{\rm a1}^*-\rm Q_{\rm a3}^*$ are implemented in an 8-mon circuit geometry,
while $\rm Q_{\rm a4}$ and  $\rm Q_{\rm a5}$  correspond to conventional X-mon circuits.
}
\label{fig:fit_flux}
	\end{center}
\end{figure}
Performing the Gaussian average yields
\begin{equation}
|W_{\rm flux}(t)|
=
\frac{1}{
\big[1+(D_2\sigma_\Phi^2 t)^2\big]^{1/4}}
\exp\!\left[
-\frac{D_1^2\sigma_\Phi^2 t^2}
{2\big(1+(D_2\sigma_\Phi^2 t)^2\big)}
\right].
\end{equation}
Away from the sweet spot this expression reduces to an approximately Gaussian decay, while near the sweet spot it exhibits a characteristic power-law behavior~\cite{RevModPhys.86.361}.
The total Ramsey envelope is given by
\begin{equation}
E(t;\Phi)
=
\exp\!\left[-\frac{\Gamma_1(\Phi)}{2}\,t\right]
\,\big|W_{\rm flux}(t;\Phi)\big|,
\end{equation}
where $\Gamma_1(\Phi)$ denotes the energy-relaxation rate.
Since the total envelope $E(t)$ is generally neither purely exponential nor Gaussian, we extract the effective Ramsey dephasing time $T_2^*$ operationally by solving $E(T_2^*)=e^{-1}$ at each flux bias point. 
This procedure does not rely on assuming any specific functional form of the decay.
To determine the optimal flux-noise amplitude $\sigma_\Phi$ that best fits the experimental data, we minimize the squared residual
\begin{equation}
\mathrm{Err}(\sigma_\Phi)
=
\sum_j
\Big[
T_{2,\mathrm{exp}}^*(\Phi_j)
-
T_{2,\mathrm{model}}^*(\Phi_j)
\Big]^2.
\end{equation}
The dependence of $\mathrm{Err}(\sigma_\Phi)$ on $\sigma_\Phi$ is further examined
to assess the robustness of the fit.

We first analyze the Ramsey dephasing under the assumption that flux noise is the sole source of pure dephasing. Using a quasi-static Gaussian flux-noise model that includes both first- and second-order flux sensitivity, we fit the experimentally measured flux-bias dependence $T_2^*$ with the flux-noise amplitude $\sigma_\Phi$ as the only free parameter.
Figure~\ref{fig:fit_flux} summarizes this analysis for multiple devices. For each device, $\sigma_\Phi$ is independently extracted by fitting the full flux-bias dependence of $T_2^*$. Device $\rm Q_{\rm a1}^*$ is shown as a representative example: the left panel displays the best fit to the experimental data, while the right panel shows the corresponding error landscape used to determine $\sigma_\Phi$. The remaining panels present the experimental data and independently optimized flux-noise--only fits for the other devices.
As can be seen from the representative device Q1, even considering the second-order flux contribution, the model that only considers flux noise is difficult to match the $T_2^*$ value at the experimentally observed optimal operating point.  Importantly, the same difference was observed in other devices as well shown in Figs~\ref{fig:fit_flux}\bt{c}-\ref{fig:fit_flux}\bt{f} , and it is not a difference unique to a single qubit.
This robust behavior indicates that additional dephasing channels, beyond quasi-static flux noise, contribute significantly near the optimal working point.

\begin{figure}[t]
\begin{center}
\includegraphics[width=14cm]{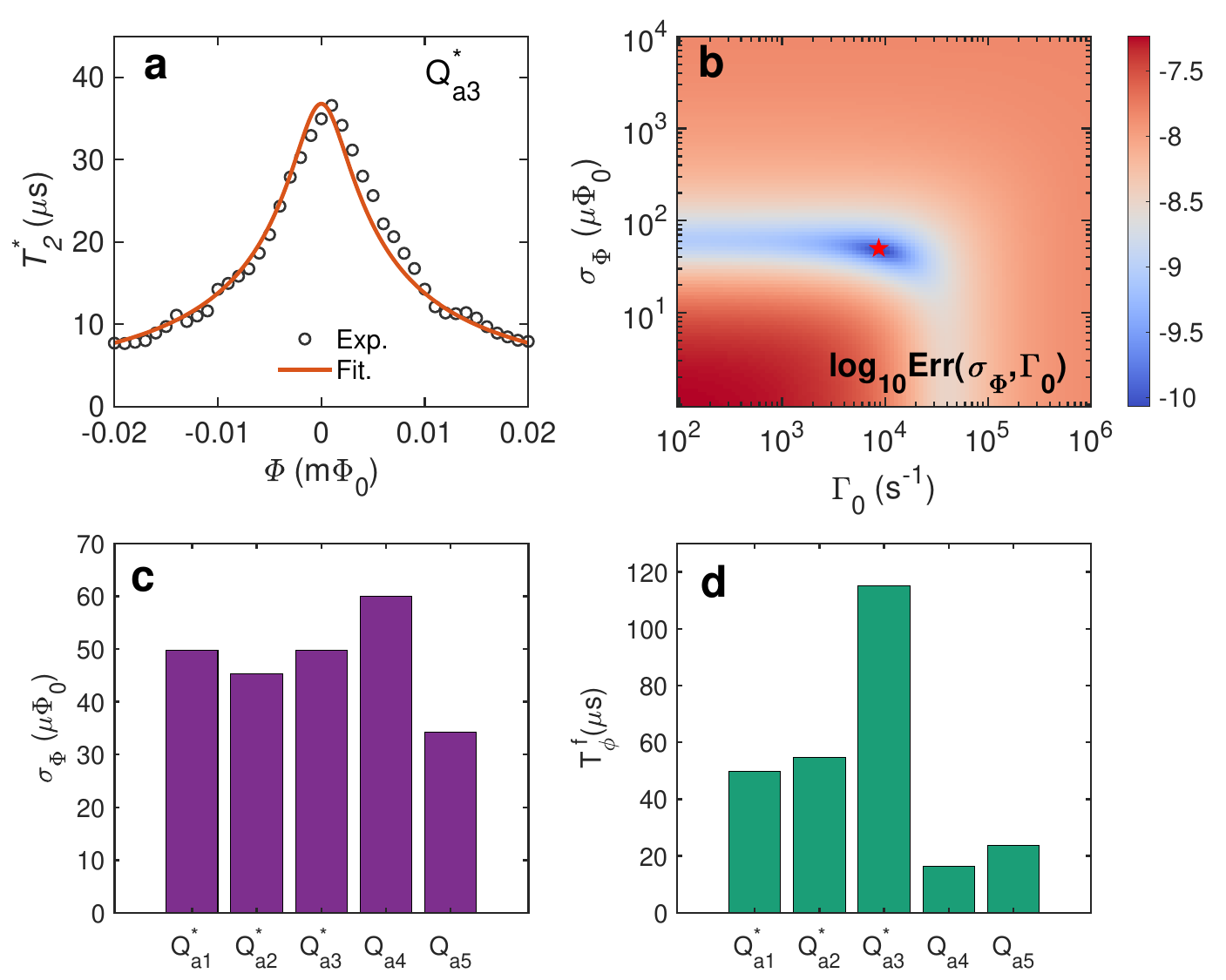}
\caption{\bt{Ramsey dephasing analysis including flux noise and  flux bias independent dephasing.}
\bt{a}, Measured Ramsey dephasing time $T_2^*$ of device $\rm Q_{\rm a3}^*$ as a function of flux bias (symbols), together with the best fit (solid line) obtained from a quasi-static flux-noise model
augmented by a  bias flux independent pure-dephasing rate $\Gamma_0$.
\bt{b}, Two-dimensional error landscape $\log_{10}\mathrm{Err}(\sigma_\Phi,\Gamma_0)$ for $\rm Q_{\rm a3}^*$, computed by fitting the full bias flux dependence of $T_2^*$.
The star marks the global minimum, which defines the optimal parameters $(\sigma_\Phi,\Gamma_0)$.
\bt{c}, Extracted effective flux-noise amplitudes $\sigma_\Phi$ for all devices, obtained from independent two-parameter fits.
\bt{d}, Corresponding flux bias  independent dephasing times
$T_\phi^{\mathrm f}=1/\Gamma_0$ for the same devices.
}
\label{fig:flux_gamma0}
	\end{center}
\end{figure}

To explain this effect, we introduce a  bias flux independent  pure-dephasing rate $\Gamma_0$.
The total Ramsey envelope is then written as
\begin{equation}
E(t;\Phi)
=
\exp\!\left[-\Big(\frac{\Gamma_1(\Phi)}{2}+\Gamma_0\Big)t\right]
\,\big|W_{\rm flux}(t;\Phi)\big|,
\end{equation}
where  $\Gamma_0$ parametrizes bias-independent dephasing processes not associated with magnetic flux fluctuations.
Figure~\ref{fig:flux_gamma0} summarizes the analysis.
Device $Q_{a3}^*$ is shown as a representative example.
Figure~\ref{fig:flux_gamma0}\bt{a} displays the measured $T_2^*$ together with the best fit obtained by simultaneously optimizing $\sigma_\Phi$ and $\Gamma_0$.
The corresponding two-dimensional error landscape is shown in
Fig.~\ref{fig:flux_gamma0}\bt{b}, where a well-defined global minimum is identified, indicating that the two parameters can be robustly extracted from the full bias flux dependence of the data.
Applying the same two-parameter fitting procedure to all devices,
we extract the effective flux-noise amplitude $\sigma_\Phi$
(Fig.~\ref{fig:flux_gamma0}\bt{c}) and the  bias flux independent dephasing time $T_\phi^{\mathrm f}=1/\Gamma_0$ (Fig.~\ref{fig:flux_gamma0}\bt{d}).
While the extracted flux-noise amplitudes vary only moderately among devices, the values of $T_\phi^{\mathrm f}$ exhibit pronounced device-to-device variation.
This behavior indicates that, although flux noise governs the overall flux-bias dependence of $T_2^*$, an additional bias flux independent dephasing channel limits the coherence time near the optimal bias point.

A natural interpretation of this bias-flux-independent dephasing contribution is that it reflects the effective influence of long-wavelength components of flux noise.
While short-wavelength flux fluctuations give rise to the pronounced bias-flux dependence of $T_2^*$ through the linear and quadratic flux sensitivities, long-wavelength fluctuations vary only weakly across the device and therefore contribute an approximately bias-independent dephasing component
within the experimentally relevant bias range.

Importantly, the parameter $\Gamma_0$ should be understood as an effective description rather than evidence for a distinct non-flux dephasing mechanism.
In particular, low-$k$ flux-noise components whose temporal statistics are not well captured by the quasi-static Gaussian approximation may give rise to an approximately exponential decay over the Ramsey time window, which is naturally absorbed into the phenomenological rate $\Gamma_0$ in our fitting procedure.
While other bias-independent dephasing mechanisms cannot be excluded solely on the basis of Ramsey measurements, interpreting $\Gamma_0$ as arising from long-range correlated flux noise provides a unified description of the observed data.
Within this picture, the extracted bias-independent dephasing time
$T_\phi^{\mathrm f}$ characterizes the effective contribution of the long-wavelength sector of flux noise, rather than indicating a separate non-flux dephasing channel.

\onecolumngrid
\clearpage
\end{document}